\documentclass{article}
\usepackage{spconf,amsmath,graphicx,hyperref}
\usepackage{booktabs}
\usepackage{cite} 

\title{EnvSSLAM-FFN: Lightweight Layer-Fused System for ESDD 2026 Challenge}
\name{
\begin{tabular}{c}
Xiaoxuan Guo\textsuperscript{1*},
Hengyan Huang\textsuperscript{1*},
Jiayi Zhou\textsuperscript{2},
Renhe Sun\textsuperscript{2},
Jian Liu\textsuperscript{2}, \\
Haonan Cheng\textsuperscript{1\dag},
Long Ye\textsuperscript{3},
Qin Zhang\textsuperscript{3}
\end{tabular}
\thanks{\dag\ Corresponding author}
\thanks{* Equal contribution}
}
\address{ 
\textsuperscript{1}State Key Laboratory  of Media Convergence and Communication, \\
Communication University of China, Beijing, China\\
\textsuperscript{2}Machine Intelligence, Ant Group, Shanghai, China\\
\textsuperscript{3}Key Laboratory of Media Audio \& Video, Ministry of Education\\
Communication University of China, Beijing, China\\
}

\begin{document}
%
\maketitle
\begin{abstract}
Recent advances in generative audio models have enabled high-fidelity environmental sound synthesis, raising serious concerns for audio security. The \emph{ESDD~2026 Challenge} therefore addresses environmental sound deepfake detection under unseen generators (Track~1) and black-box low-resource detection (Track~2) conditions. We propose \emph{EnvSSLAM-FFN}, which integrates a frozen SSLAM self-supervised encoder with a lightweight FFN back-end. To effectively capture spoofing artifacts under severe data imbalance, we fuse intermediate SSLAM representations from layers~4--9 and adopt a class-weighted training objective. Experimental results show that  the proposed system consistently outperforms the official baselines on both tracks, achieving Test Equal Error Rates (EERs) of 1.20\% and 1.05\%, respectively.
\end{abstract}

\section{Introduction}
\label{sec:intro}
Recent advances in generative audio models (e.g., AudioLDM~\cite{liu2023audioldm}, AudioGen~\cite{kreuk2023audiogen}) have enabled the synthesis of high-fidelity environmental sounds, posing significant threats to audio security. Unlike structured speech, environmental audio exhibits vast spectral and temporal variability, causing detectors trained on speech-centric self-supervised models~\cite{chen2022wavlm,hsu2021hubert} and benchmarks like ASVspoof~\cite{nautsch2021asvspoof2019,liu2023asvspoof2021} or ADD~\cite{yi2022add} to struggle with generalization. The \emph{ESDD~2026 Challenge}~\cite{yin2025esdd} addresses these challenges by evaluating environmental deepfake detection under unseen generators (Track 1) and black-box low-resource detection (Track 2) conditions.

In this paper, we propose \emph{EnvSSLAM-FFN}, a lightweight system that integrates a frozen SSLAM~\cite{alex2025sslam} encoder with a FFN back-end. To better characterize spoofing artifacts under severe data imbalance, the system fuses intermediate SSLAM representations from layers~4--9 and adopts a class-weighted training objective. This design emphasizes task-relevant acoustic cues for environmental sound forensics. Experimental results show that \emph{EnvSSLAM-FFN} consistently outperforms the official baselines, achieving Test EERs of 1.20\% on Track~1 and 1.05\% on Track~2, respectively.

\begin{figure*}[t]
    \centering
    \includegraphics[width=0.8\linewidth]{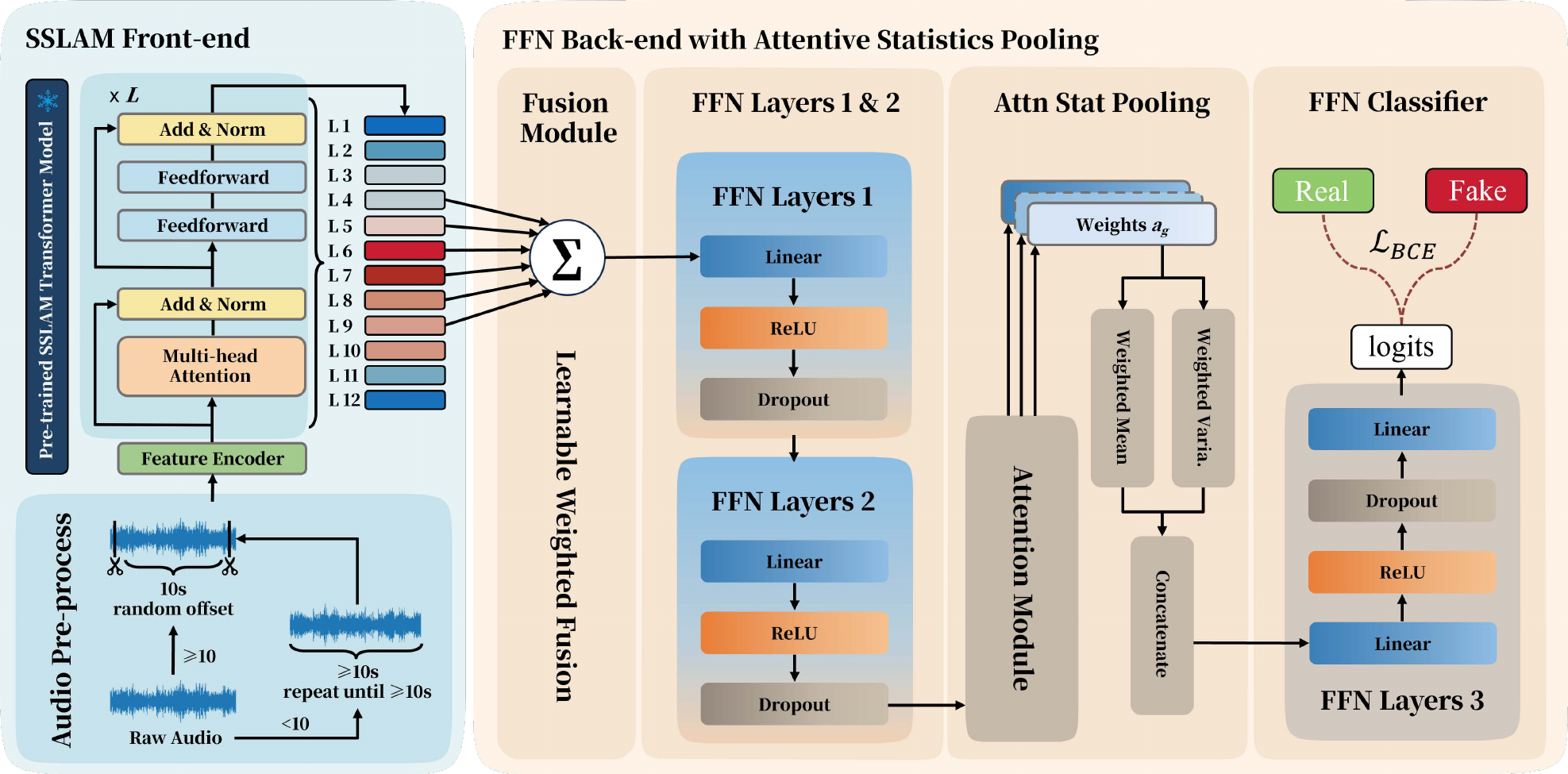}
    \caption{Overview of the proposed EnvSSLAM-FFN pipeline}
    \label{fig:framework}
\end{figure*}

\begin{figure}[t]
    \centering
    \begin{minipage}{1\linewidth}
        \includegraphics[width=\linewidth]{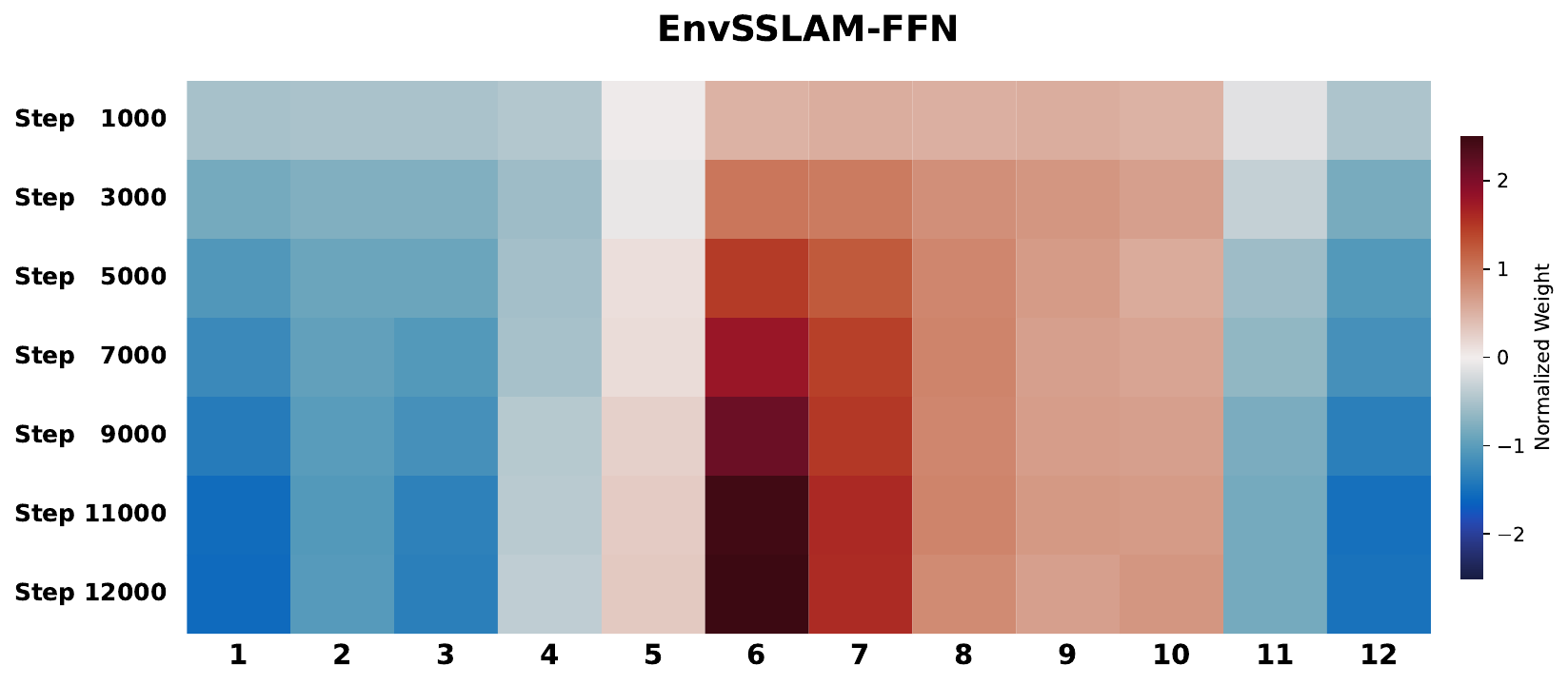}
    \end{minipage}
    \caption{Trends of normalized fusion weights across the 12 SSLAM layers in EnvSSLAM‑FFN over training steps.}
    \label{fig:hot}
\end{figure}

\section{Proposed System}
\label{sec:system}
\noindent \textbf{Data Processing.}
All raw recordings are resampled to 16\,kHz and converted to mono. To ensure temporal consistency while enabling data augmentation, clips shorter than 10\,s are repeated, and longer ones are truncated to 10\,s using a random offset strategy. We extract 128-dimensional log-Mel filter-bank features (HTK-compatible, Hanning window, 10\,ms shift) to match the input requirements of SSLAM~\cite{alex2025sslam}. The features are normalized using global statistics and shaped to a fixed length of 1{,}024 frames.

\smallskip
\noindent\textbf{Model Architecture.}
Fig.~\ref{fig:framework} illustrates the overall \emph{EnvSSLAM-FFN} pipeline, where a frozen SSLAM front-end extracts frame-level embeddings that are subsequently processed by a lightweight FFN back-end~\cite{elkheir2025comprehensive} with two projection layers using ReLU activation and dropout.~Temporal information is aggregated via attentive statistics pooling~\cite{okabe2018attentive}, which computes weighted means and standard deviations over the frame sequence, and the pooled representation is fed to a shared classifier producing bona-fide and spoof scores for both tracks. To mitigate the severe label imbalance, training employs a class-weighted binary cross-entropy loss, assigning higher weights to bona-fide samples according to the inverse class-frequency ratio.

\smallskip
\noindent \textbf{Layer Fusion.}
We employ a learnable fusion module that aggregates SSLAM hidden states using softmax weights. As illustrated in Fig.~\ref{fig:hot}, the fusion weights gradually concentrate on the middle layers (approximately layers~4--9) during training, while both shallow and deep layers receive lower emphasis. This observation indicates that intermediate SSLAM representations provide more informative cues for anti-spoofing, compared with low-level acoustic patterns in early layers and semantic-oriented features in deeper layers. \emph{EnvSSLAM-FFN} therefore restricts the fusion to layers~4--9 in the final system.

\begin{table}[t]
    \centering
    \caption{EER (\%) on Track~1 and Track~2 for the official baselines and our EnvSSLAM-FFN system.}
    \label{tab:track_results}
    \setlength{\tabcolsep}{6pt}
    \begin{tabular}{c c c c}
        \toprule
        \# & System for Track 1 & Eval EER & Test EER \\
        \midrule
        1 & AASIST (baseline)        & 15.26 & 15.02 \\
        2 & BEATs+AASIST (baseline)  & 14.21 & 13.20 \\
        \midrule
        3 & \textbf{EnvSSLAM-FFN (ours)} & \textbf{1.05} & \textbf{1.20} \\
        \midrule\midrule
        \# & System for Track 2 & Eval EER & Test EER \\
        \midrule
        1 & AASIST (baseline)        & 15.72 & 15.40 \\
        2 & BEATs+AASIST (baseline)  & 12.64 & 12.48 \\
        \midrule
        3 & \textbf{EnvSSLAM-FFN (ours)} & \textbf{1.24} & \textbf{1.05} \\
        \bottomrule
    \end{tabular}
\end{table}

\section{Experiments}
\label{sec:experiments}
\noindent \textbf{Experimental Setup.} 
We evaluate the proposed system on the EnvSDD dataset~\cite{yin2025esdd}, which contains 45.25 hours of bona-fide audio and 316.7 hours of synthetic data. For Track~1, the training set includes 27{,}811 real samples and 111{,}244 synthetic deepfake samples generated by multiple generators, and we optimize the FFN back-end utilizing Adam~\cite{kingma2014adam} with a learning rate of $10^{-4}$, batch size 32, and dropout 0.1, selecting the best checkpoint via development-set EER. Track~2 provides 270 real and 1{,}083 black-box spoofed samples for low-resource adaptation; the system is initialized with the best Track 1 weights and fine-tuned on the Track 2 training split with a reduced learning rate of $5\times10^{-5}$. Both tracks employ the class-weighted objective to mitigate label imbalance.

\smallskip
\noindent \textbf{Experimental Results.} 
Table \ref{tab:track_results} presents the EER results for both tracks. In Track 1 (unseen generators), \emph{EnvSSLAM-FFN} reduces the Eval EER from 14.21\% (BEATs+AASIST) to 1.05\%, while achieves a Test EER of 1.20\%, representing a significant improvement over the 13.20\% baseline. In Track 2 (black-box low-resource detection), our system lowers the Eval EER from 12.64\% to 1.24\% and the Test EER from 12.48\% to 1.05\%. These results show that intermediate SSLAM fusion, combined with lightweight fine-tuning, enables effective detection and adaptation in both tracks.

\section{Conclusion}
\label{sec:conclusion}
We proposed EnvSSLAM-FFN, integrating a frozen SSLAM encoder with a FFN back-end for the ESDD 2026 Challenge. Leveraging intermediate layer fusion and class-weighted training, our system significantly outperforms official baselines on both tracks. We achieve Test EERs of 1.20\% and 1.05\% on Track 1 and Track 2, respectively.

\section*{Acknowledgment}
This work was supported by the Beijing Natural Science Foundation (No.~4252011) and the National Natural Science Foundation of China (No.~62271455).
\bibliographystyle{IEEEbib}
\bibliography{strings,refs}

@misc{yin2025esdd,
  author = {Han Yin and Yang Xiao and Rohan Kumar Das and Jisheng Bai and Ting Dang},
  title = {ESDD 2026: Environmental Sound Deepfake Detection Challenge Evaluation Plan},
  howpublished = {arXiv preprint arXiv:2508.04529},
  year = {2025},
  note = {preprint}
}

@inproceedings{elkheir2025comprehensive,
  title     = {Comprehensive Layer-wise Analysis of SSL Models for Audio Deepfake Detection},
  author    = {El Kheir, Yassine and Samih, Younes and Maharjan, Suraj and Polzehl, Tim and M{\"o}ller, Sebastian},
  booktitle = {Findings of the Association for Computational Linguistics: NAACL 2025},
  pages     = {4070--4082},
  year      = {2025}
}

@misc{alex2025sslam,
  author = {Tony Alex and Sara Ahmed and Armin Mustafa and Muhammad Awais and Philip J. B. Jackson},
  title = {SSLAM: Enhancing Self-Supervised Models with Audio Mixtures for Polyphonic Soundscapes},
  howpublished = {arXiv preprint arXiv:2506.12222},
  year = {2025},
  note = {preprint}
}

@article{nautsch2021asvspoof2019,
  title        = {ASVspoof 2019: Spoofing Countermeasures for the Detection of Synthesized, Converted and Replayed Speech},
  author       = {Nautsch, Andreas and Wang, Xin and Evans, Nicholas and Kinnunen, Tomi H. and Vestman, Ville and Todisco, Massimiliano and Delgado, H{\'e}ctor and Sahidullah, Md and Yamagishi, Junichi and Lee, Kong Aik},
  journal      = {IEEE Transactions on Biometrics, Behavior, and Identity Science},
  volume       = {3},
  number       = {2},
  pages        = {252--265},
  year         = {2021}
}

@article{liu2023asvspoof2021,
  title        = {ASVspoof 2021: Towards Spoofed and Deepfake Speech Detection in the Wild},
  author       = {Liu, Xuechen and Wang, Xin and Sahidullah, Md and Patino, Jose and Delgado, H{\'e}ctor and Kinnunen, Tomi and Todisco, Massimiliano and others},
  journal      = {IEEE/ACM Transactions on Audio, Speech, and Language Processing},
  volume       = {31},
  pages        = {2507--2522},
  year         = {2023}
}

@inproceedings{yi2022add,
  title        = {ADD 2022: The First Audio Deep Synthesis Detection Challenge},
  author       = {Yi, Jiangyan and Fu, Ruibo and Tao, Jianhua and Nie, Shuai and Ma, Haoxin and Wang, Chenglong and Wang, Tao and others},
  booktitle    = {ICASSP 2022 -- 2022 IEEE International Conference on Acoustics, Speech and Signal Processing (ICASSP)},
  pages        = {9216--9220},
  year         = {2022},
  organization = {IEEE}
}

@article{liu2023audioldm,
  title={AudioLDM: Text-to-Audio Generation with Latent Diffusion Models},
  author={Liu, Haohe and Chen, Zehua and Yuan, Yi and Mei, Xinhao and Liu, Xubo and Mandic, Danilo and Wang, Wenwu and Plumbley, Mark D},
  journal={arXiv preprint arXiv:2301.12503},
  year={2023}
}

@inproceedings{kreuk2023audiogen,
  title={AudioGen: Texturally Guided Audio Generation},
  author={Kreuk, Felix and Synnaeve, Gabriel and Polyak, Adam and Singer, Uriel and Alexandre, D{\'e}fossez and Copet, Jade and Parikh, Devi and Taigman, Yaniv and Adi, Yossi},
  booktitle={International Conference on Learning Representations (ICLR)},
  year={2023}
}

@article{chen2022wavlm,
  title={WavLM: Large-Scale Self-Supervised Pre-Training for Full Stack Speech Processing},
  author={Chen, Sanyuan and Wang, Chenanyi and Chen, Zhengyang and Wu, Yu and Liu, Shujie and Chen, Zhuo and Li, Jinyu and Kanda, Naoyuki and Yoshioka, Takuya and Xiao, Xiong and others},
  journal={IEEE Journal of Selected Topics in Signal Processing},
  volume={16},
  number={6},
  pages={1505--1518},
  year={2022},
  publisher={IEEE}
}

@article{hsu2021hubert,
  title={HuBERT: Self-Supervised Speech Representation Learning by Masked Prediction of Hidden Units},
  author={Hsu, Wei-Ning and Bolte, Benjamin and Tsai, Yao-Hung Hubert and Lakhotia, Kushal and Salakhutdinov, Ruslan and Mohamed, Abdelrahman},
  journal={IEEE/ACM Transactions on Audio, Speech, and Language Processing},
  volume={29},
  pages={3451--3460},
  year={2021},
  publisher={IEEE}
}

@inproceedings{okabe2018attentive,
  title={Attentive Statistics Pooling for Deep Speaker Embedding},
  author={Okabe, Koji and Koshinaka, Takafumi and Shinoda, Koichi},
  booktitle={Proc. Interspeech},
  pages={2252--2256},
  year={2018}
}

@article{kingma2014adam,
  title={Adam: A Method for Stochastic Optimization},
  author={Kingma, Diederik P and Ba, Jimmy},
  journal={arXiv preprint arXiv:1412.6980},
  year={2014}
}
\end{document}